\documentclass[12pt]{iopart}
\begin{document}

\title{Cosmological consequences of a built-in cosmological
constant model}

\author{Arbab I. Arbab$^{1,2}$\footnote[1]{arbab@ictp.trieste.it}}

\address{ $^1$ Department of Physics, Teacher's
College, Riyadh 11491, P.O.Box 4341, Kingdom of Saudi Arabia.}

\address{ $^2$ Comboni College for Computer Science, P.O. Box 114, 
Khartoum, Sudan.}

\begin{abstract}
We have analyzed the implications of a phenomenological model of a built-in
cosmological constant of  Rastall and Al-Rawaf -Taha type. We have produced 
the
presently observed cosmic acceleration with the aid of introducing a 
positive
cosmological constant. We have shown that the
 cosmological models of the type $\Lambda\propto\frac{\ddot R}{R},
 \Lambda\propto H^2$ or $\Lambda\propto 8\pi G\rho$ are equivalent
 to a built-in cosmological constant model of Al-Rawaf and Taha type.
Such models are compatible with the recent observational data and do not 
suffer from
fine tuning problems. We predict a higher lensing probability than in 
Einstein de
Sitter  and Abdel Rahman models. We also predict a higher value for the 
luminosity
and angular diameter distances than that of Abdel Rahman model. The lens 
distribution
peaks at relatively smaller values of the lens red-shift and occurs at 
higher
red-shift.

\end{abstract}


\maketitle

\section{Introduction}
Gravitational lensing is currently known as one of  the robust
tests of cosmological models with variable cosmological constant
($\Lambda$) (Fukugita {\it et al.} 1992, Caroll {\it et al} 1992).
It is found by Ratra and Quillen (Ratra and Quillen 1998) that the
optical depth is significantly smaller than that in a constant
$\Lambda$ model with the same density parameter ($\Omega$). They
also found that the red-shift of the maximum of the lens
distribution falls between that in the constant-$\Lambda$
model and the Einstein-de Sitter model (ES). \\
It is recently shown that our present universe is accelerating.
This apparent acceleration is attributed to a dark energy residing
in space itself, which also balances the kinetic energy of
expansion so as to give the universe zero spatial curvature, as
deduced from the cosmic microwave background radiation (CMBR).
This dark energy was important in the past as it is now. It might
have played a part in limiting the formation of largest
gravitationally bound structures. One can model the dark energy by
a cosmological constant (or a vacuum decaying energy). However,
not all vacuum decaying cosmological models predict this
acceleration. We have recently proposed a cosmological model with
a cosmological constant of the form
$\Lambda=\beta\left(\frac{\ddot R}{R}\right)$, where $R$ is the
scale factor of the universe and $\beta$ some constant. We have
found for $\Lambda>0$ cosmic acceleration is inevitable. This
cosmological constant mimics a cosmic fluid (scalar field) with an
equation of state of the form $p_\phi=\omega_\phi \rho_\phi $
($-1<\omega_\phi<-1/3$). Such a field is some times known as {\it
quintessence}. Quintessence models exhibit an event horizon which
poses a serious problem for string theory (Fichler 2001).
Quintessence models describe the present expansion without resort
to the cosmological constant. The question whether this
acceleration will last forever or changes is not yet known!   If
the present vacuum contribution ($\Omega_\Lambda)$ is such that
$\Omega_\Lambda=\frac{2}{3}$ then this cosmological constant
mimics a string-like matter. Present observations indicate that
the matter contribution ($\Omega_m$) to the present universe is
such that $0.1<\Omega_m<0.4$. One possibility is that this
$\Lambda$ may be an effective constant, and indeed could be
varying slowly with time. Such a behavior can be modelled by the
potential energy of a scalar field, in exactly the same way as
inflation is.
\\
In this work, we show the similarity between a phenomenological
models of a type of $\Lambda\propto H^2$ and $\Lambda\propto
\frac{\ddot R}{R}$ and the one with $\Lambda \propto 8\pi G\rho$.
These forms of $\Lambda$ are also equivalent to the Rastall
(Rastall 1972) and Al-Rawaf and Taha (Al-Rawaf and Taha 1996)
models of modified general relativity (MGR). Al Rawaf and Taha
obtained a built-in cosmological constant that is related to the
curvature scalar ($\cal{R}$). In MGR a flat universe has an age
($t_0$) of $t_0<H^{-1}_0$, where $H_0$ is the present value of the
Hubble constant. However, our model gives a resolution to both the
age problem and the low energy density problem of the universe
with present observational data. We also discuss the cosmological
tests pertaining to diameter-distance, luminosity-distance and
gravitational lensing effect. It is now known that the
gravitational lensing probability is quite sensitive to the value
of $\Lambda$ in low-$\Omega$, flat models. It has also been noted
that for a source at a given red-shift, the red-shift of the peak
of the expected normalized probability distribution for the lens
red-shift  depends on the value of $\Lambda$. We have found that
our present model gives rather higher values for the luminosity
and angular diameter distances with the presently preferred value
for $\Omega_m (=0.3)$. The constraints implied by these tests
would place a limit on the value of $\beta$.
\section{The Field Equations}

The Einstein field equations with a cosmological constant, and
energy conservation law yield \begin{equation} \left(\frac{\dot
R}{R}\right)^2+\frac{k}{R^2}=\frac{8\pi}{3}
G\rho+\frac{\Lambda}{3} \end{equation} \begin{equation}
\frac{\ddot R}{R}=-\frac{4\pi G}{3}(\rho+3p)+\frac{\Lambda}{3}
\end{equation} and
\begin{equation}
\dot\rho+3\frac{\dot R}{R}( p+\rho)=-\frac{\dot\Lambda}{8\pi G}.
\end{equation}
Using the equation of the state
\begin{equation}
p=(\gamma-1)\rho, \qquad \ 1\le \gamma \le 2 \ ,\end{equation}
eq.(2) can be written as
\begin{equation} \frac{\ddot R}{R}=\frac{8\pi
G}{3}\left(1-\frac{3}{2}\gamma\right)\rho +\frac{\Lambda}{3}.
\end{equation} Following (Arbab 2002), we consider
\begin{equation} \Lambda=\beta\left(\frac{\ddot R}{R}\right),
\end{equation}
where $\beta$ is constant. From eqs.(1)-(5), one finds (Arbab
2002) for a flat universe ($k=0$) \begin{equation} R(t)=C
t^{\frac{\beta-2}{\beta-3}},\
 \qquad C=\rm const.\ , \ \beta\ne 3\ \end{equation}
\begin{equation}
\Lambda(t)=\frac{\beta(\beta-2)}{(\beta-3)^2}\frac{1}{t^2}\ ,
\qquad
 \beta\ne 3, \end{equation}
 \begin{equation} \rho(t)=\frac{(\beta-2)}{(\beta-3)}\frac{1}{4\pi
Gt^2}\ \ , \qquad \beta\ne 3\end{equation} the deceleration
parameter is given by
\begin{equation} q=-\frac{\ddot RR}{\dot
R^2}=\left(\frac{1}{2-\beta}\right), \qquad \ \ \beta\ne 2
\end{equation} The matter contribution to density parameter of the
universe is
\begin{equation} \Omega_m=\frac{2}{3}\frac{(\beta-3)}{(\beta-2)}\
, \qquad \ \beta\ne 2 \end{equation} while the vacuum contribution
is
\begin{equation} \Omega_\Lambda=\frac{\beta}{3(\beta-2)}, \qquad \
\beta\ne 2, \end{equation} so that $\Omega_m+\Omega_\Lambda=1$, as
preferred by inflation. From eqs.(11) and (12) one finds
\begin{equation}
\beta=\frac{2\Omega_\Lambda}{\Omega_\Lambda-\frac{1}{3}}=\frac{1-\Omega_m}
{\frac{1}{3}
-\frac{1}{2}\Omega_m}\ . \end{equation} It is thus clear that the
free parameter ($\beta$) can be determined from the present
observational data from the knowledge of the vacuum or baryonic
mass contribution to the total energy density of the universe. For
$\beta >0$  eq.(13) gives $\Omega_\Lambda > \frac{1}{3}$ (or
$\Omega_m < \frac{2}{3}$). We see that $\beta=6$,
$\Omega_m=\Omega_\Lambda=\frac{1}{2}$. For $\beta >6$, $\Omega_m
>\Omega_\Lambda$ and for $\beta<6$, $\Omega_m <\Omega_\Lambda$. Note that 
all
other cosmological parameters depend on this constant. The case
$\beta=2$ defines a static universe, which is physically
unacceptable  for describing  the present universe . We see that
the universe will be ultimately  driven into a de-Sitter phase of
exponential expansion as $\Omega_\Lambda\rightarrow 1$ ( or $\beta
\rightarrow 3$) (Johri 2000,
Arbab 2002). \\
 As suggested by many people, that $\Omega_\Lambda=\frac{2}{3}$
(or $\Omega_m=\frac{1}{3}$), one would obtain the following
constraints: \begin{equation} \beta=4,\qquad \ H_0t_0=2\ , \qquad
\ q_0=-0.5\ . \end{equation} These findings have to be confronted
with current observation. Thus, if the universe is dominated by
vacuum, today, then it should accelerate. Note that RT model does
not exhibit an acceleration. This because a cosmic acceleration
would mean a negative matter density parameter. This why RT
predicts a shorter age ($H_0t_0=0.85$) of the universe, unless
Hubble constant assumes a smaller value.

We would like to remark that our cosmological constant is always
positive ($\Lambda>0$) irrespective of the sign of $\beta$. One
only  has a decelerating universe ($q>0$) when $\beta<2$.
\section{A cosmological constant and matter content}

In general, the most important difference of a dark energy
component to a cosmological constant is that its equation of state
can be different form $p=-\rho$, generally implying a
time-variation. Using eqs.(4) and (5) one can write eq.(6) as
\begin{equation}
\Lambda=\left(\frac{\beta}{3-\beta}\right)\left(1-\frac{3}{2}\gamma\right)8
\pi
G\rho \ ,\qquad \ \beta\ne 3 .
\end{equation}
It is evident that when $\gamma=\frac{2}{3}$ the cosmological
constant vanishes ($\Lambda=0$). Thus if the universe is dominated
by strings today, the cosmological constant must vanish! Such a
model is obtained by Sima and Sukenik (Sima and Sukenik 2001) for
a non-decelerative
universe. \\
For the matter-dominated universe ($\gamma=1$) eq.(15)
gives\begin{equation} \Lambda^{\rm
AR}=\left(\frac{\beta}{\beta-3}\right)4\pi G\rho^m\ .
\end{equation} It is evident from the above equation that an empty
universe ($\rho=0$) would imply a vanishing cosmological constant 
($\Lambda=0$). \\
For the radiation-dominated epoch ($\gamma=\frac{4}{3}$) eq.(15)
gives
\begin{equation} \Lambda=\left(\frac{\beta}{\beta-3}\right)8\pi
G\rho^r\ . \end{equation} If $\beta$ changes then eqs. (16) and
(17) can be written as defining $\beta$ (with
$\rho_v=\frac{\Lambda}{8\pi G}$) as
\begin{equation} \beta^m=\frac{6\rho^m_{v}}{2\rho^m_{v}-\rho^m}\
 , \qquad \beta^r=\frac{3\rho^r_{v}}{\rho^r_{v}-\rho^r}\ ,
\end{equation} where the subscripts ``r"  and ``m" denote the
value of the quantity during radiation and matter epochs,
respectively. We see that for $\beta >0$, the vacuum contribution
always surpasses the ordinary matter/radiation contribution to the
total energy density of the universe in both radiation era
($\rho^r_v > \rho^r$) and matter era ($ \rho^m_v >
\frac{1}{2}\rho^m$). Therefore, the vacuum energy (cosmological
constant) has played a very essential role in inflating the
universe in the past, and accelerating it in the present epoch.
Thus without the cosmological constant the observable universe
would not have been produced. One then should not worry about the
fine tuning problem infected other cosmologies inasmuch as the
ratio between the vacuum energy to ordinary matter/radiation
energy does not change with time.
\section{A built-in cosmological constant}
The curvature scalar $\cal{R}$ is given by
\begin{equation}
{\cal{R}}=6\left[\left(\frac{\dot R}{R}\right)^2+\frac{\ddot
R}{R}\right].
\end{equation}
Using eqs.(1), (5) and (15) the above equation reads
\begin{equation}
\Lambda^{\rm AR}=\frac{\beta}{6(\beta-1)}\cal{R}.
\end{equation}
 Recently, Al-Rawaf and Taha (RT), in an attempt to solve the entropy 
problem,
proposed a theory in which the energy conservation law is relaxed.
They have shown that
 \begin{equation}
 \Lambda^{\rm RT}=\frac{(1-\eta)}{3(2-\eta)}\cal{R}, \end{equation}
where $0\le\eta\le 1$ and \begin{equation} \Omega^{\rm RT}_m=\eta\
,\ \qquad  \eta=2q\ .\end{equation} Similarly Majernik (Majernik
2001, 2002) proposed a cosmological model with the ansatz
\begin{equation}
\Lambda^{\rm MJ}=8\pi \kappa GT
\end{equation}
where $T$ is the stress-energy scalar ($T^\mu_\mu $) and $\kappa$
is some constant. And since we know that $T$ is related to
$\cal{R}$,  one can write for Majernik (MJ)
\begin{equation}
\Lambda^{\rm MJ}=\frac{\kappa}{1+4\kappa}\cal{R}.
\end{equation}
Thus eqs.(20), (21) and (24) are different types of a built-in
cosmological constant. Though the three models are similar in the
matter-dominated era (MD) they differ in both eras.\\
(i) In the radiation- dominated era (RD) eqs.(21) and
(24) yield $\Lambda=0$ whereas eq.(20) does not. \\
(ii) In the matter-dominated era eqs.(20) and (24) lead to cosmic
acceleration of the universe whereas  eq.(21) does not. \\ We see
that our model interplays between the two. Comparison of eq.(20)
with eq.(21) using eqs.(10) and (11) yields
\begin{equation} \Omega_m=\frac{2}{3}+\frac{\eta}{3}\ \ , \qquad \
\eta=2q=\frac{2}{2-\beta}. \end{equation} We note that, unlike RT,
our model can solve both the age  and the low mass problems of the
universe simultaneously. Our model gives
\begin{equation} t_0=\frac{2}{3}\frac{H_0^{-1}}{\Omega_m}
\end{equation}
while RT yields
\begin{equation}
t_0=\frac{2H_0^{-1}}{2+\Omega_m}.
\end{equation}
In RT the age of the universe is limited by the condition
$H_0t_0<1$, while in our model $H_0t_0\ge 1$, in a good agreement
with the present observational data. This is an advantage over RT
model, where the age problem and the low-mass problem can not be
solved simultaneously. We remark that while our model can account
for the presently accelerating universe, RT can not accommodate
this possibility, as this is evident from eq.(25).

Substituting eq.(14) in eq.(2) one yields
\begin{equation}
\frac{\ddot R}{R}=\frac{4\pi G}{3}\frac{3}{(\beta-3)}\rho\ .
\end{equation}
For a positive energy density ($\rho >0$), as seen  from eq.(9),
$\beta >3$. This implies that $\ddot R >0$ ( not with
$p<-\frac{1}{3}\rho$, as usual). Consequently, one obtains a new
cosmic acceleration that has not been explored before. Hence, the
observed acceleration, as indicated by supernovae of type Ia, may
not require an exotic equation of state, as suggested by many
scientists. A physically acceptable solution requires $\beta
>3$ and this represents the robust constraint for our model.
However, this automatically give rise to cosmic acceleration.
Hence cosmic acceleration could have started at any time whenever
$\frac{1}{3}<\Omega_\Lambda<1$.

A similar setting, as in eq.(28), is recently suggested by Gu and
Hang (Gu and Hang 2001). However, we have found that this would
imply $p=-\rho$. Therefore, ordinary matter in the presence of a
positive cosmological constant can render cosmic acceleration.
However, in Al-Rawaf and Taha the factor $\frac{3}{(3-\beta)}$ in
eq.(16) is absorbed in the definition of the gravitational
constant $G$ making their field equations different from that of
Einstein. This is done by replacing $G$ by [ $G/\alpha$ ] (or
equivalently $G\rightarrow G/(\frac{3}{3-\beta}$)). However,
Al-Rawaf and Taha can describe an accelerating universe (with the
normal
equation of state) if the gravitational constant is negative! \\
In what follows we discuss the following cosmological tests:
\section{Neoclassical tests}
A photon emitted by a source with co-ordinate $r=r_1$ at time
$t=t_1$ and received at time $t_0$ by an observer located at $r=0$
will follow a null geodesic. The proper distance between the
source and the observer is given by
\begin{equation}
d(z)=R_0\int_R^{R_0}\frac{dR}{R\dot R} \end{equation} From
eq.(7)one obtains [Arbab 1998]
\begin{equation}
H_0d=\frac{1}{(\frac{3}{2}\Omega_m-1)}\left[1-(1+z)^{(1-\frac{3}{2}
\Omega_m)}\right],
\end{equation}
where $1+z=\frac{R_0}{R}$ defines the red-shift $z$.
\subsection{Luminosity distance} This is given by
\begin{equation}
d_L=\left(\frac{L}{4\pi\ell}\right)^{1/2}=r_1R_0(1+z)=d(z)(1+z)\ ,
\end{equation}
where $L$ is the total power emitted by the source and $\ell$ is
the apparent luminosity of the object at a distance $r_1$. Using
eq.(30) one gets
\begin{equation}
H_0d_L=\frac{1+z}{(\frac{3}{2}\Omega_m-1)}\left[1-(1+z)^{(1-\frac{3}{2}
\Omega_m)}\right]
\end{equation}
We see that $H_0d_L$ is a decreasing function of $\Omega_m$.
$H_0d_L$ {\it versus} $z$ is plotted in fig.(1) for $\Omega_m=0.3$
for AM, ES and our model (AR). We observe that our model gives a
higher value for $H_0d_L$ than AM and ES.

\subsection{Angular diameter distance} The angular diameter ($d_A$)
of a light source of proper distance $d$ is given by
\begin{equation}
d_A=\frac{d(z)}{(1+z)}\ ,
\end{equation}
or
\begin{equation}
d_AH_0=\frac{1}{(\frac{3}{2}\Omega_m-1)}\left[\frac{1-(1+z)^{(1-\frac{3}{2}
\Omega_m)}}{1+z}\right].
\end{equation}
This has a maximum at a red-shift ($z_m$) given by
\begin{equation}
1+z_m=\left(\frac{3}{2}\Omega_m
\right)^{1/(\frac{3}{2}\Omega_m-1)}
\end{equation}
Again $H_0d_A$ is a decreasing function of $\Omega_m$. $H_0d_A$
{\it versus} $z$ is plotted in fig.(2) for $\Omega_m=0.3$ for AM,
ES and our model (AR). The maximum red-shift in our model (AR)
occurs at $z_m=0.76$.

\subsection{Gravitational Lensing} Following Abdel Rahman
solution, one finds that the lensing probability equation for the
present  model is given by (Fukugita, {\it et al.} 1992; Caroll
{\it et al.} 1992; Cohn 1992)
\begin{equation}
P_{lens}=\frac{1}{8}\left[\frac{(1-x_s^{(\frac{3}{2}\Omega_m-1)})}{(\frac{3}
{2}\Omega_m-1)(1-x_s^{1/2})}\right]^3
, \end{equation} the normalized optical depth for lensing is given
by (Ratra and Quillen 1992)
\begin{equation}
\tau(z_s)=\frac{1}{30}\left[\frac{(1-x_s^{(\frac{3}{2}\Omega_m-1)})}{(\frac
{3}{2}\Omega_m-1)}\right]^3
\end{equation}
and the normalized lens red-shift distribution is given by
\begin{equation}
\tau^{-1}\frac{d\tau}{dz}=F\left(30\
x^{\Omega_m/(\frac{3}{2}\Omega_m-1)}\right),
\end{equation}
\begin{equation} F=\left(\frac{(1-x^{(\frac{3}{2}\Omega_m-1)})^2(x^{(\frac
{3}{2}\Omega_m-1)}
-x_s^{(\frac{3}{2}\Omega_m-1)})^2}{
(\frac{3}{2}\Omega_m-1)(1-x_s^{(\frac{3}{2}\Omega_m-1)})^5}\right),
\end{equation}
where $x=(1+z)^{-1}$ and $z_s$ is the source red-shift. In Abdel
Rahman model $\eta$ replaces $\frac{2}{2-\beta }$ and both models
coincide with the Einstein-de Sitter model (ES) when $\eta=1$ (or
$\beta=0$), but differ otherwise. We plot the lensing probability
(fig.(3)), the normalized optical depth (fig.(4)) and the
normalized lens red-shift distribution (fig.(5)) against the
source red-shift ($z_s$) and matter density ($\Omega_m$) for Abdel
Rahman (AM) model and our present model (AR). We have found that
in our model the normalized lens red-shift distribution peaks at
relatively  smaller values of the lens red-shift than that of AM
and ES, and occurs at relatively high values of the lens
red-shift. The lensing probability in our model is very high for a
very low matter universe.
\section{Discussion}
All in all, since higher values for $\Omega_\Lambda$ predict
higher numbers of gravitational lenses, this technique offers a
viable way of putting upper bounds on the value of the
cosmological constant. The normalized lens red-shift
 distribution for $\Omega_m=0.3$ peaks at  smaller values in our model in 
comparison
 with AM,  and occurs at a higher lens red-shift. For a low density universe
  the optical  depth rises very quickly with the source red-shift than in 
AM model.
  The lens distribution  peaks at considerably lower values of the lens red-
shift than in AM.
 The diameter and luminosity distances are  decreasing functions of 
$\Omega_m$.
 Our model predicts  rather higher values for  these cosmological
 distances.We predict far more lens system for a low-density universe than 
might have
 been observed.
 We have shown that a  cosmological model of the type $\Lambda\propto\frac
{\ddot R}{R},
 \Lambda\propto H^2$ or $\Lambda\propto 8\pi G\rho$ are equivalent
 to a built-in cosmological constant model of Al-Rawaf and Taha
 type. Different cosmological tests are worked out and some results
 are shown in fig.(1) to fig(5). The ensuing future results can
 limit the presently different cosmological scenarios to a few ones.
\subsection*{Acknowledgements}
I would like to express my thanks and appreciations to the abdus
Salam ICTP (Trieste) for hospitality and the Swedish Foundation
for Science and Technology (SIDA) for financial support.
\subsection*{ References}
Abdel Rahman, A.-M.,  Astrophys. Space Sci.278, 383(2001) \\
Al-Rawaf, A.S., and Taha, M.O.,  Gen. Rel. Gravit.28, 935(1996)\\
Arbab, A.I., Class. Quantum Grav. 20, 93 (2003)\\
Fukugita, {\it et al.},  Astrophys J. 393, 99 (1992)
Gu, J. and Hwang, W., arXiv.gov/abs/astro-ph/0106387\\
Fichler, W., {\it et al}., arXiv.gov/abs/hep-th/0104181\\
Johri, V.B., arXiv.gov/abs/astro-ph/0007079\\
Majern$\rm \acute{i}k$, V., Phys. Lett. A282, 362(2001), gr-qc/0201019\\
Rastall, P,  Phys. Rev.D6, 3357(1972)\\
Ratra, B., and Quillen, A.,  MNRAS 259, 738(1992)\\
Riess, A.G., {\it et al.},  Ap. J.116, 1009(1998)\\
Sima, J., and  Sukenik, M. arxiv.org/abs/gr-qc/0105090
\\  Hellerman, S., {\it et
\label{lastpage}
\end{document}

-----------------------------
Dr. Arbab I. Arbab, CPhys. MInst.P
Department of Physics,
Teacher's  College,
Riyadh, Postal Code 11491.
P.O. Box 4341,
Saudi Arabia

Mobile: 0096657451544 
Te. 0096612336183